\documentclass[11pt]{article}
\usepackage{graphicx} 
\usepackage[english]{babel}
\usepackage[affil-it]{authblk} 
\usepackage{etoolbox}
\usepackage{lmodern}
\DeclareUnicodeCharacter{2212}{-} 
\usepackage[letterpaper,top=3cm,bottom=3cm,left=3cm,right=3cm,marginparwidth=1.75cm]{geometry}
\usepackage{amsmath}
\usepackage{graphicx}
\usepackage[colorlinks=true, allcolors=blue]{hyperref}

\begin{document}

\title{Ferromagnetic resonance in Y$_3$AlFe$_4$O$_{12}$ garnets}

\author[1]{D.~Popadiuk}
\author[2]{V.~Borynskyi}
\author[1,2]{A.~Kravets}
\author[3]{Y.~Shlapa}
\author[3]{S.~Solopan}
\author[3]{A.~ Belous}
\author[2]{A.~Tovstolytkin}
\author[1]{V.~Korenivski}

\affil[1]{\footnotesize Nanostructure Physics, Royal Institute of Technology, 10691 Stockholm, Sweden}
\affil[2]{\footnotesize V.G. Baryakhtar Institute of Magnetism of the NAS of Ukraine, 03142 Kyiv, Ukraine}
\affil[3]{\footnotesize V.I. Vernadsky Institute of General and Inorganic Chemistry of the NAS of Ukraine, Kyiv 03142, Ukraine}

\date{}

\maketitle

\begin{abstract}
Spin dynamics in Al-substituted yttrium iron garnets is investigated using broadband ferromagnetic resonance measurements in the temperature range $T=200\div 360$~K. Using the measured data, the resonance field and linewidth as well as their temperature dependence are determined, with implications for the uniformity and overall quality of the samples prepared via different chemical fabrication routes. These key parameters governing the spin dynamics in the material are important for its applications in high-speed spintronic and magnonic devices.
\end{abstract}

\section*{Introduction}

Advances in modern high-speed information and communication technologies motivate intensive research efforts on magnetic oxides, whose vanishing conductivity hence low eddy currents losses make them the materials of choice in microwave applications~\cite{dionne2009magnetic,harris2011modern,ZHENG2022162471,Kefeni2017,6678201}. Yttrium iron garnet (YIG, Y$_3$Fe$_5$O$_{12}$) is particularly suited for microwave electronics, spintronics, magnonics, optoelectronics, owing to its record low spin damping, controllable saturation magnetization, high electrical resistivity, and large magneto-optical Faraday rotation~\cite{YANG2021158235,10.1063/5.0044993,Serga_2010,Wu2013}. YIGs magnetic properties can be controlled via various fabrication routes~\cite{Serga_2010, Schmidt20207, 10.1063/5.0047054}, though tuning the material for a specific operating frequency and temperature can be challenging~\cite{Dionne19975064,harris2022modern,PENAGARCIA20205871,WU2021157996}. 

The physical properties of YIGs strongly depend on their chemical composition, fabrication details, and the resultant microstructure~\cite{LI2018936,Cheng2008,MUSA20171135}. Partial substitutions in the cationic sub-lattices was found to be particularly effective~\cite{Cheng2008,BASAVAD202012015}. Thus, Al-doped YIG (Y$_3$AlFe$_4$O$_{12}$) ferrites have shown low dielectric and magnetic losses as well as higher dielectric and magnetic permittivity when compared to the base YIG ferrite~\cite{KIM2003553}. Our own recent results~\cite{BORYNSKYI2025178320,shlapa2024yttrium} showed improved static magnetic properties, such as near-zero coercivity at room temperature, promising for low operating fields and improved switching performance in magneto-electronic devices. Seldom tested is the temperature stability and uniformity of the spin-dynamic parameters, important for developing YIG-based materials since microwave ferrites often need to operate in a wide temperature range.

This work investigates the spin dynamics in the Al-substituted YIG ferrites fabricated using the method of precipitation from aqueous solutions that we recently developed~\cite{BELOUS2021158140, SOLOPAN2023172248}. For this we use broadband ferromagnetic resonance (FMR) measurements over a temperature range encompassing the technologically most important room temperature (RT). We analyze the FMR-extracted temperature-dependent parameters of the material, such as the resonance field and resonance linewidth (spin damping), which govern the microwave response of the material,  important for its applications.

\section*{Methods and characterization}

The samples in the form of ceramic tablets were obtained by sintering at 1350-1400~deg.~C of Y$_3$AlFe$_4$O$_{12}$ nanopowder synthesized via co-precipitation from an aqueous solution as described in detail in~\cite{BELOUS2021158140,BORYNSKYI2025178320,shlapa2024yttrium,SOLOPAN2023172248}. The crystallographic properties of the synthesized Y$_3$AlFe$_4$O$_{12}$ garnets were studied using a DRON-4 X-ray diffractometer (XRD) equipped with a CuK$_\alpha$ radiation tube. The microstructure of the samples was analyzed using a JEM 1230 and JEM 1400 transmission electron microscopes (Jeol, Japan) and a FEG-SEM Nova Nanosem 230 FEI and SEC miniSEM SNE4500 MB scanning electron microscopes. 

The processes occurring at the precipitation stage are of particular importance since they can strongly affect the properties of the resulting powders. Three approaches to precipitation were used for comparison~\cite{BELOUS2021158140,shlapa2024yttrium}. Stoichiometric amounts of metal salts were precipitated simultaneously using NaOH solution at the constant pH of the medium $8.8\div 8.9$ (sample S1); consecutive precipitation using the solution of NaOH, with precipitation of the mixture of Fe(OH)$_3$ and Al(OH)$_3$ hydroxides at pH $4\div 4.5$ followed by that of the Y(OH)$_3$ hydroxide at pH $8.8\div 8.9$ (sample S2); consecutive precipitation using the solution of NH$_4$OH, with precipitation of the mixture of Fe(OH)$_3$ and Al(OH)$_3$ hydroxides at pH $4\div 4.5$ followed by that of the Y(OH)$_3$ hydroxide at pH $8.8\div 8.9$ (sample S3).

The samples for ferromagnetic resonance (FMR) studies were cut in the form of rectangular parallelepipeds with dimensions 5×3×1 mm$^3$. The FMR measurements were carried out using a DynaCool PPMS system (Quantum Design Inc.) equipped with a custom-made broadband coplanar waveguide setup. The FMR spectra were rerecorded at fixed temperature ($T=200\div 360$~K) and frequency ($f=$7, 9, and 12~GHz) by sweeping the magnetic field in the range zero to 5~kG (9~Tesla range available in the PPMS). After each sweep at a given $T$ and $f$, the field was reduced to zero and the frequency or sample temperature was stepped before the next field sweep.

All three synthesized ceramic samples were determined by XRD to be single-phase. The crystal structure belongs to the Ia-3d~\cite{Wu2013} cubic space group with a lattice constant of 1.2321(5)~nm. The value of the lattice constant is smaller than that of the pure YIG (1.2376~nm) and consistent with the values reported for Y$_3$AlFe$_4$O$_{12}$ in Refs.~\cite{AZADIMOTLAGH20091980,Musa2017}. The measured density for the YAIG ceramic samples was 4.96~g/cm$^3$, which is only 1.4\% less than the calculated via XRD data (5.03~g/cm$^3$). This indicataces that our samples have low porosity. The average grain size was determined to be $\approx 10$ $\mathrm{\mu m}$.

\section*{Results and discussion}

\begin{figure}[!ht]
    \centering
    \includegraphics[width=1\linewidth]{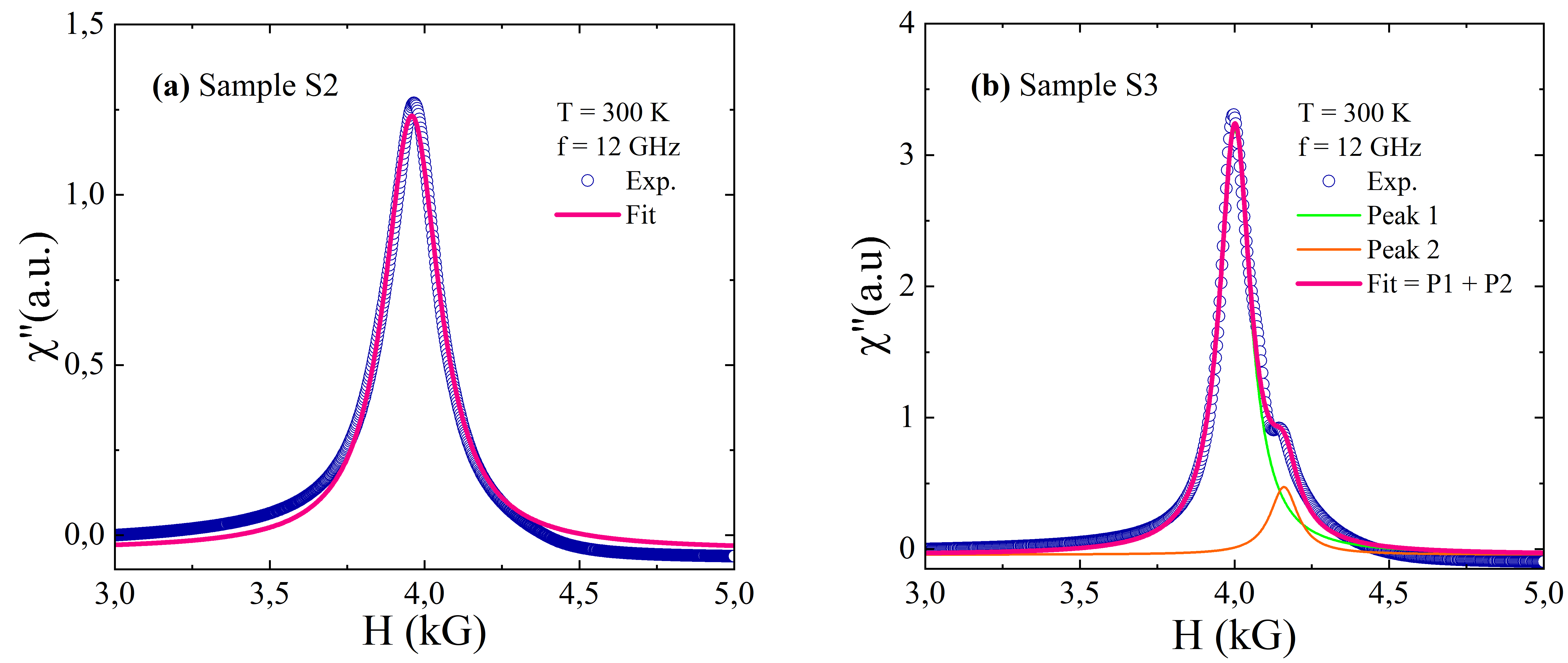}
    \caption{Examples of FMR spectra (integrated microwave signal) recorded at 12~GHz and 300~K, with fitting to extract FMR parameters, for: (a) sample S2, with one FMR peak yielding good fit to data; (b) sampleS3, with two sub-resonances clearly seen in fitting.}
    \label{fig:spectra}
\end{figure}

Broadband ferromagnetic resonance (FMR) measurements were performed at various temperatures using a coplanar waveguide excited by a signal from a radio-frequency generator ($0.01\div 12$~GHz range). The RF signal was rectified by a microwave diode and fed into a lockin amplifier synced to a low-field modulation coil surrounding the sample-waveguide fixture (operating at low $f\sim 10$~Hz). The external DC field ($H$) was applied along the longest side of the samples. The RF field excitation ($\sim 1$~Oe in amplitude) was applied perpendicular to the DC field. The fixed-frequency field spectra of the FMR measurement circuit, proportional to the field-derivative of the imaginary part of the magnetic susceptibility of the sample (a measure of the magnetic dissipation in the studied material), were recorded for a given temperature and integrated to obtain $\chi''(H)$. Two such spectra for $T=300$~K and $f=12$~GHz are shown in Fig.~\ref{fig:spectra} for samples S2 and S3 (with S1 being similar in FMR properties to S3, not shown).

Though our previously reported magnetostatic magnetometry measurements showed single $M-H$ loops~\cite{BORYNSKYI2025178320,shlapa2024yttrium}, the FMR results are able to finely differentiate the samples as to their single-phase character. Thus, sample S2 shows a single FMR peak well fitted by a single Lorentzian, as shown in Fig.~\ref{fig:spectra}(a). In contrast, samples S1and S3 showed a signature of a second resonance (not uncommon for other similar materials~\cite{Barron2017,Verma2024}), clearly distinguishable by fitting the data with a combination of two Lorentzian peaks. This is illustrated in Fig.~\ref{fig:spectra}(b) for S3, where good fits to the experimental data could only be obtained by using one pronounced and one satellite resonance in each case.

\begin{figure}[!ht]
    \centering
\includegraphics[width=1\linewidth]{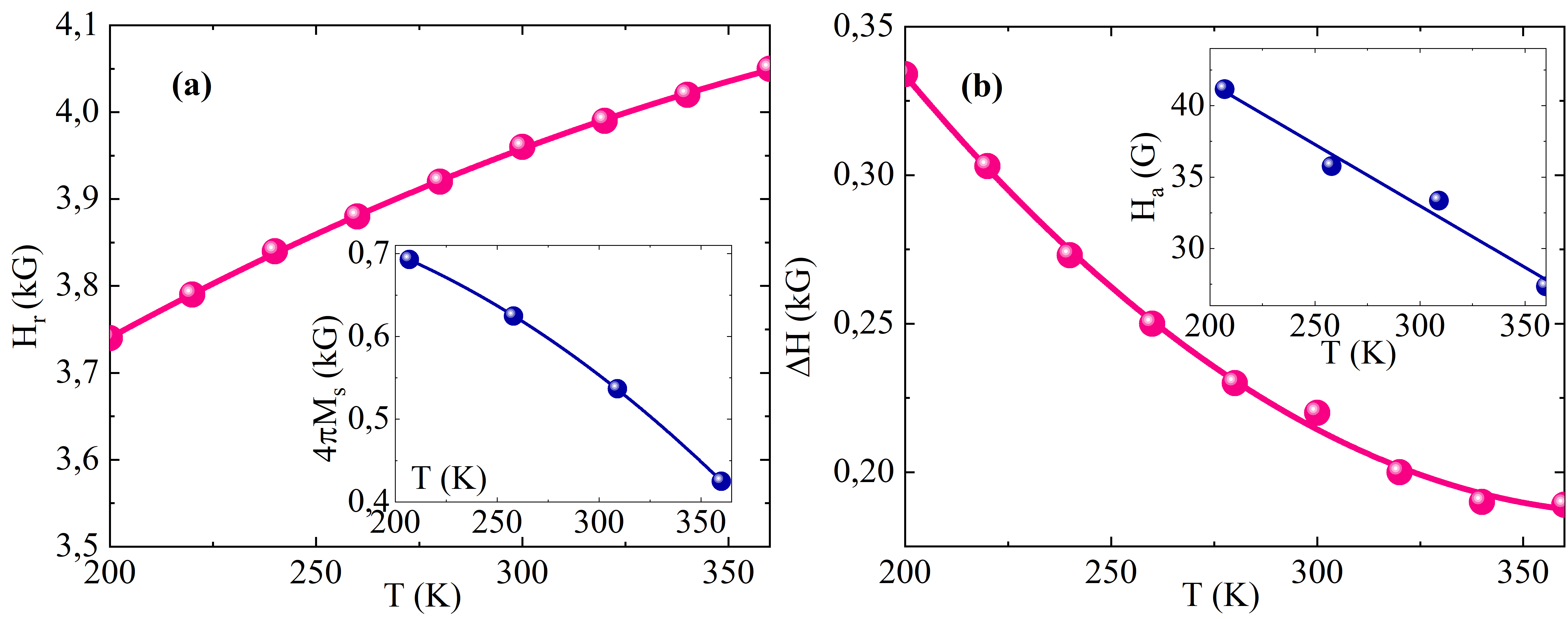}
    \caption{Temperature dependence of FMR resonance field $H_\mathrm{r}$ (a) and linewidth $\Delta H$ (b) for sample S2, $f=12$~GHz. Insets show corresponding magnetization $4\pi M_\mathrm{s}(T)$ and anisotropy field $H_\mathrm{a}$ adapted from Ref.~\cite{BORYNSKYI2025178320} for $T=200\div 360$~K.}
    \label{fig:Hr-dH}
\end{figure}

The resonance field versus temperature $H_\mathrm{r}(T)$ obtained from fitting the FMR spectra for sample S2 is shown in Fig.~\ref{fig:Hr-dH}(a). As expected, $H_\mathrm{r}$ decreases somewhat with decreasing $T$, owing primarily to the rising magnetization at low temperature, as shown in the inset. The original $M(T)$ data from Ref.~\cite{BORYNSKYI2025178320} for $T=200\div 360$~K was adapted to reflect the temperature dependence of the respective contribution to the effective field (demag-fraction of $4\pi M$ in Gauss). Even though the samples are bulk rather than thin films, their shape is far from spherical, the smallest ratio of the sides is 3:1 and, therefore, a significant fraction of the magnetization 
is acting as a demag-field. More specifically, using the demagnetizing factors for a rectangular prism~\cite{aharoni1998demagnetizing} (in our case with the sides $x:y:z=1:3:5$~mm, magnetized along the length, $z$), $N_x^D=0.6372$, $N_y^D=0.2287$, $N_z^D=0.134$, and the change in the magnetization in the measurement temperature range $\Delta M_\mathrm{s}(200\div 360\mathrm{K})\approx 300$~G, the change in the relevant shape anisotropy field contributing to the effective FRM field is estimated to be approximately $4\pi \Delta M_\mathrm{s}(N_x^D-N_y^D)\approx 100\div150$~G. 
This is much larger than the change in the measured anisotropy field in this temperature range ($\sim10$~G, see discussion below). 

Measurements at 7, 9, and 12~GHz (at RT, for sample S2) gave the resonance fields of, respectively, 2.21, 2.86, and 3.97~kG, which are slightly below the simple Larmor estimates for zero demag-FMR (for would be ideally spherical samples)~\cite{Ozgur2009}. This confirms the above estimate of the demag-contribution being of the order of 100~G, susceptible to a small thermal variation due to varying $M(T)$. 

Fig.~\ref{fig:Hr-dH}(b) shows the full width at half maximum (FWHM) as a function of temperature extracted from the FMR spectra of the type shown in Fig.~\ref{fig:spectra}, measured at $f=12$~GHz for S2. The determination of the resonance width for this sample is straightforward as the FMR peak is fitted best by a single Lorentzian. $\Delta H$ characterizes the strength of the spin dissipation in the material, also known as spin damping, directly proportional to the Gilbert damping constant $\alpha$. To maintain a steady state precession at the FMR, energy must be supplied by the microwave field since the precession is subject to several microscopic dissipation processes~\cite{kittel2018introduction,cullity2011introduction}. This can be magnon-magnon scattering with excitation of higher-order spin-wave modes and diminishing uniform FMR, which can be significant for some materials and for some specific sample shapes. Omnipresent, however, is magnon-lattice dissipation via the spin-orbit interaction found in all magnetic crystals. The same spin-orbit coupling is responsible for the magnetic anisotropy, so the magnitude of the effective anisotropy field $H_\mathrm{a}$ can be considered as a measure of the strength of the spin-orbit coupling and, hence, of the spin dissipation via the spin-orbit channel. The inset to Fig.~\ref{fig:Hr-dH}(b) shows our $H_\mathrm{a}(T)$ data from Ref.~\cite{BORYNSKYI2025178320} for $T=200\div 360$~K, which is adapted to reflect qualitatively the expected temperature dependence of the spin-orbit contribution to $\Delta H(T)$ extracted from the measured FMR spectra. As seen from the data in the inset, the change in $H_\mathrm{a}$ over the experimental temperature interval is $\sim 10$~G, which is an order of magnitude smaller than the estimated change in the demag-field from $M(T)$ contributing to the effective FMR field. We note that the phonon population does not change drastically over the temperature range in question. 

\begin{figure}[!ht]
    \centering
    \includegraphics[width=0.48\linewidth]{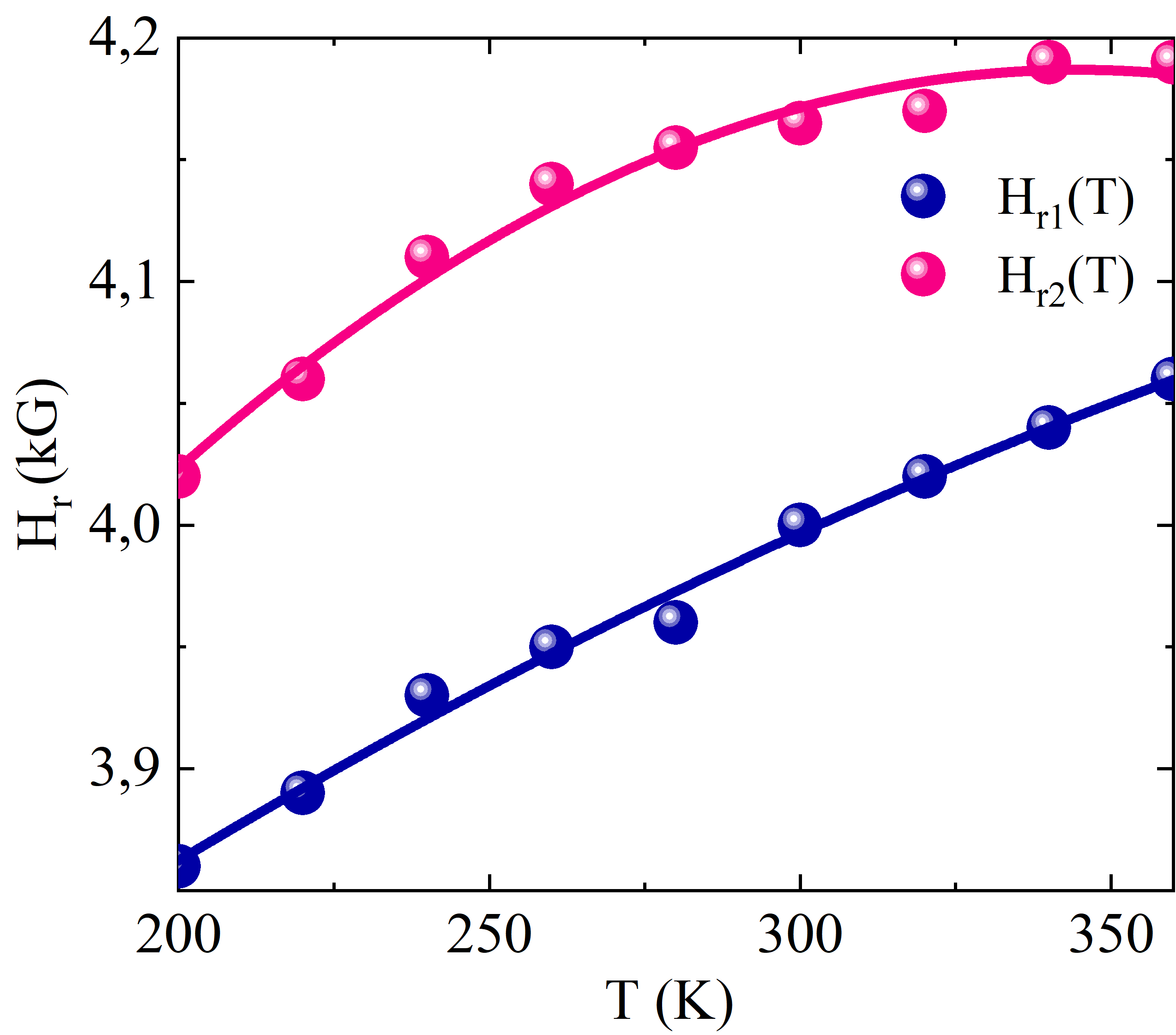}
    \caption{Two resonance fields extracted from fitting FMR spectra for sample S3 vs. temperature. Double-FMR peak persists at all experimental temperatures.}
    \label{fig:Hr1Hr2-T}
\end{figure}

The slight inhomogeneity visible for sample S3 (and S1) only in the FMR data (not in $M(H,T)$) as a weak satellite resonance peak persists at all temperatures, as shown in Fig.~\ref{fig:Hr1Hr2-T}. Even though our detailed XRD studies show single-phase material for all samples, it is natural to expect that in ceramics sintered from nano-grained powder there is some variation in properties between the crystallites and their grain boundaries. Such variations were not detected in magnetometry~\cite{BORYNSKYI2025178320,shlapa2024yttrium}, which uniformly shows single $M-H$ loops. FMR thus proves to be a higher sensitivity technique when it come to detecting fine variations in the homogeneity of the materials, in this case reflected potentially in variations of both the magnetization magnitude and the local anisotropy field of the inter-grain interfaces. 

Although a contribution from paramagnetic ions to the measured spectra cannot be excluded in principle, the respective electron paramagnetic resonance (EPR) signal is expected to be negligible. Indeed, even a hypothetical clustering of non-magnetic Al leading to an isolated paramagnetic impurity (e.g., Fe$3^+$ surrounded predominantly by Al neighbors, strongly suppressing the on-site exchange interaction) would still have the effective paramagnetic volume fraction at $\sim 10^{-2}$ level at most (much less for any realistic inhomogeneity in this material) simply because of the small Al proportion in the compound (1:7 versus the magnetic ions Fe and Y). Perhaps even more importantly, the experimental temperature and field ranges used of about 300~K and 0.3~T should lead to a 1000-fold thermally induced reduction in the presumed EPR signal as the latter is known to scale with the Boltzmann factor, $\mu_BH/k_BT\sim 10^{-3}$ in this case. In contrast, the FMR absorbed power is proportional to $M^2$ in a ferromagnet, which is a big fraction of $M_\mathrm{s}$ at RT. An EPR signal of $\sim 10^{-5}$ (at most) on the scale of the measured FMR signal would be undetectable in our experiment.

It should be interesting to investigate the lower temperature region, below 150~K, where large changes in the magnetization and anisotropy were found~\cite{BORYNSKYI2025178320}, to see how these changes correlate with the resonance properties observed in the differently fabricated samples. As an example, if the slight splitting of the FMR would be the effect of a weakened exchange for the surface (interface) spins having otherwise same magnitude, the two resonance peaks could shift closer or merge at near zero temperature. 

\section*{Conclusions}

The observed changes in the FMR resonance field $H_\mathrm{r}$ and linewidth $\Delta H$ with temperature are overall consistent with our previously reported magnetostatic data for these Y$_3$AlFe$_4$O$_{12}$ materials~\cite{BORYNSKYI2025178320,shlapa2024yttrium}, specifically with the changes observed in the magnetization and anisotropy with temperature. Sample S2 prepared by separately precipitating Y$^{3+}$ cations (delayed with respect to Fe$^{3+}$ and Al$^{3+}$) using sodium hydroxide is found to have the best FMR characteristics, with a single, rather narrow resonance in the measured temperature range. Other samples show a small inhomogeneity, with a satellite FMR peak persisting at all studied temperatures. The results presented underline the importance of FMR spectroscopy as a sensitive characterization technique, unique in the context of microwave applications, and illuminate additional aspects of the studied Y$_3$AlFe$_4$O$_{12}$ garnets important for further work on optimizing the fabrication process. 

\section*{Acknowledgements}
Support from the Swedish Research Council (VR 2018-03526), Olle Engkvist Foundation (2020-207-0460), Swedish Strategic Research Council (SSF UKR24-0002) are gratefully acknowledged. This work was partially supported by the IEEE Magnetics Society Program (Magnetism for Ukraine – 2023) (STCU Project No. 9918) and NAS of Ukraine (grant No 0124U002212 in the framework of the Target Program “Grants of the NAS of Ukraine to research laboratories/groups of young scientists of the NAS of Ukraine” (2024–2025)). V. B. is grateful for the support from the NRFU grant nr. 2020.02/0261.

\bibliographystyle{unsrt}
\bibliography{refs.bib}

\begin{thebibliography}{10}

\bibitem{dionne2009magnetic}
G~F Dionne.
\newblock {\em Magnetic oxides}, volume~14.
\newblock Springer, 2009.

\bibitem{harris2011modern}
V~G Harris.
\newblock Modern microwave ferrites.
\newblock {\em IEEE Transactions on Magnetics}, 48:1075, 2011.

\bibitem{ZHENG2022162471}
Zongliang Zheng, Xu~Wu, Fengjiao Li, Ping Yin, and Vincent~G. Harris.
\newblock Low-loss $\mathrm{NiZnCo}$ ferrite composites with tunable magneto-dielectric performances for high-frequency applications.
\newblock {\em Journal of Alloys and Compounds}, 894:162471, 2022.

\bibitem{Kefeni2017}
K.~K. Kefeni, T.~A.~M. Msagati, and B.~B. Mamba.
\newblock Ferrite nanoparticles: synthesis, characterisation and applications in electronic device.
\newblock {\em Materials Science and Engineering B}, 215:37, 2017.

\bibitem{6678201}
Andrew~D. Block, Prabesh Dulal, Bethanie J.~H. Stadler, and Nicholas C.~A. Seaton.
\newblock Growth parameters of fully crystallized $\mathrm{YIG}$, $\mathrm{BiYIG}$, and $\mathrm{CeYIG}$ films with high faraday rotations.
\newblock {\em IEEE Photonics Journal}, 6:1, 2014.

\bibitem{YANG2021158235}
Yucong Yang, Tao Liu, Lei Bi, and Longjiang Deng.
\newblock Recent advances in development of magnetic garnet thin films for applications in spintronics and photonics.
\newblock {\em Journal of Alloys and Compounds}, 860:158235, 2021.

\bibitem{10.1063/5.0044993}
José~Diogo Costa, Bruno Figeys, Xiao Sun, Nele Van~Hoovels, Harrie A.~C. Tilmans, Florin Ciubotaru, and Christoph Adelmann.
\newblock Compact tunable $\mathrm{YIG}$-based rf resonators.
\newblock {\em Applied Physics Letters}, 118:162406, 2021.

\bibitem{Serga_2010}
A~A Serga, A~V Chumak, and B~Hillebrands.
\newblock $\mathrm{YIG}$ magnonics.
\newblock {\em Journal of Physics D: Applied Physics}, 43:264002, 2010.

\bibitem{Wu2013}
M.~Wu and A.~Hoffmann.
\newblock {\em Recent Advances in Magnetic Insulators - from Spintronics to Microwave Applications}.
\newblock Elsevier Science \& Technology Books, 2013.

\bibitem{Schmidt20207}
Georg Schmidt, Christoph Hauser, Philip Trempler, Maximilian Paleschke, and Evangelos~Th. Papaioannou.
\newblock Ultra thin films of yttrium iron garnet with very low damping: A review.
\newblock {\em Physica Status Solidi (b)}, 257:1900644, 2020.

\bibitem{10.1063/5.0047054}
Yi~Li, Chenbo Zhao, Wei Zhang, Axel Hoffmann, and Valentyn Novosad.
\newblock Advances in coherent coupling between magnons and acoustic phonons.
\newblock {\em APL Materials}, 9:060902, 2021.

\bibitem{Dionne19975064}
Gerald~F. Dionne.
\newblock Properties of ferrites at low temperatures (invited).
\newblock {\em Journal of Applied Physics}, 81:5064, 1997.

\bibitem{harris2022modern}
Vincent~G Harris.
\newblock {\em Modern Ferrites, Volume 2: Emerging Technologies and Applications}.
\newblock John Wiley \& Sons, 2022.

\bibitem{PENAGARCIA20205871}
R.~Peña-Garcia, Y.~Guerra, D.M. Oliveira, A.~Franco, and E.~Padrón-Hernández.
\newblock Local atomic disorder and temperature dependence of saturation magnetization in yttrium iron garnet.
\newblock {\em Ceramics International}, 46:5871, 2020.

\bibitem{WU2021157996}
Huarui Wu, Fengzhen Huang, Ruixia Ti, Xiaomei Lu, Chenyang Zhang, Long Yuan, Yan Xu, and Liwei Zhang.
\newblock Effect of $\mathrm{Ca}$ dopant on magnetic and magnetodielectric properties of $\mathrm{Y}_3 \mathrm{Fe}_5 \mathrm{O}_{12}$.
\newblock {\em Journal of Alloys and Compounds}, 861:157996, 2021.

\bibitem{LI2018936}
Haiyan Li and Yuheng Guo.
\newblock High microwave absorption characteristic nanomaterial preparation and mechanism analysis.
\newblock {\em Journal of Alloys and Compounds}, 765:936, 2018.

\bibitem{Cheng2008}
Zhongjun Cheng, Hua Yang, Lianxiang Yu, and Wei Xu.
\newblock Saturation magnetic properties of $\mathrm{Y}_{3-x} \mathrm{Re}_x \mathrm{Fe}_5 \mathrm{O}_{12}$ ($\mathrm{Re: Gd, Dy, Nd, Sm, La}$) nanoparticles grown by a sol--gel method.
\newblock {\em Journal of Materials Science: Materials in Electronics}, 19:442, 2008.

\bibitem{MUSA20171135}
Makiyyu~Abdullahi Musa, Raba'ah~Syahidah Azis, Nurul~Huda Osman, Jumiah Hassan, and Tasiu Zangina.
\newblock Structural and magnetic properties of yttrium iron garnet ($\mathrm{YIG}$) and yttrium aluminum iron garnet ($\mathrm{YAIG}$) nanoferrite via sol-gel synthesis.
\newblock {\em Results in Physics}, 7:1135, 2017.

\bibitem{BASAVAD202012015}
M.~Basavad, H.~Shokrollahi, H.~Ahmadvand, and S.M. Arab.
\newblock Structural, magnetic and magneto-optical properties of the bulk and thin film synthesized cerium- and praseodymium-doped yttrium iron garnet.
\newblock {\em Ceramics International}, 46:12015, 2020.

\bibitem{KIM2003553}
Chul~Sung Kim, Byoung~Ki Min, Sam~Jin Kim, Sung~Ro Yoon, and Young~Rang Uhm.
\newblock Crystallographic and magnetic properties of $\mathrm{Y}_3 \mathrm{Fe}_{5-x} \mathrm{Al}_x \mathrm{O}_{12}$.
\newblock {\em Journal of Magnetism and Magnetic Materials}, 254-255:553, 2003.
\newblock Proceedings of the 15th International Conference on Soft Magnetic Materials (SMM15).

\bibitem{BORYNSKYI2025178320}
Vladyslav Borynskyi, Dariia Popadiuk, Anatolii Kravets, Yuliia Shlapa, Serhii Solopan, Vladislav Korenivski, Anatolii Belous, and Alexandr Tovstolytkin.
\newblock Room- and low-temperature magnetic parameters of $\mathrm{Y}_3 \mathrm{AlFe}_4 \mathrm{O}_{12}$ garnets.
\newblock {\em Journal of Alloys and Compounds}, 1010:178320, 2025.

\bibitem{shlapa2024yttrium}
Yuliia Shlapa, Serhii Solopan, Vladyslav Borynskyi, Dariia Popadiuk, Anatolii Kravets, Vladislav Korenivski, Alexandr Tovstolytkin, and Anatolii Belous.
\newblock Yttrium-iron ferrite garnets: Structural and magnetic properties vs precipitation route.
\newblock In {\em 2024 IEEE 14th International Conference Nanomaterials: Applications \& Properties}. IEEE, 2024.

\bibitem{BELOUS2021158140}
A.~Belous, A.~Tovstolytkin, O.~Fedorchuk, Y.~Shlapa, S.~Solopan, and B.~Khomenko.
\newblock Al-doped yttrium iron garnets $\mathrm{Y}_3 \mathrm{AlFe}_4 \mathrm{O}_{12}$: Synthesis and properties.
\newblock {\em Journal of Alloys and Compounds}, 856:158140, 2021.

\bibitem{SOLOPAN2023172248}
S.~Solopan, A.~Tovstolytkin, V.~Zamorskyi, Yu. Shlapa, V.-A. Maraloiu, O.~Fedorchuk, and A.~Belous.
\newblock Nanoscale $\mathrm{Y}_3 \mathrm{AlFe}_4 \mathrm{O}_{12}$ garnets: Looking into subtle features of crystalline structure and properties formation.
\newblock {\em Journal of Alloys and Compounds}, 968:172248, 2023.

\bibitem{AZADIMOTLAGH20091980}
Z.~{Azadi Motlagh}, M.~Mozaffari, and J.~Amighian.
\newblock Preparation of nano-sized al-substituted yttrium iron garnets by the mechanochemical method and investigation of their magnetic properties.
\newblock {\em Journal of Magnetism and Magnetic Materials}, 321:1980, 2009.

\bibitem{Musa2017}
M.~A. Musa, R.~S. Azis, N.~H. Osman, J.~Hassan, and M.~M. Dihom.
\newblock {\em Structural and magnetic properties of yttrium aluminum iron garnet ($\mathrm{YAIG}$) nanoferrite prepared via auto-combustion sol–gel synthesis}, volume~54.
\newblock 2017.

\bibitem{Barron2017}
J.~F. Barrón, H.~Montiel, V.~Gómez-Vidales, A.~Conde-Gallardo, and G.~Alvarez.
\newblock Yig films through synthesis by means of the polymeric precursor method: Correlation between the structural and vibrational properties with magnetic behavior.
\newblock {\em Journal of Superconductivity and Novel Magnetism}, 30:2515, 2017.

\bibitem{Verma2024}
Sachin Verma, Manjushree Maity, Abhishek Maurya, Rajeev Singh, and Biswanath Bhoi.
\newblock Evolution of microstructure, magnetic and microwave properties of sputter deposited polycrystalline $\mathrm{YIG}$ thin films.
\newblock {\em Journal of Materials Science: Materials in Electronics}, 35:105, 2024.

\bibitem{aharoni1998demagnetizing}
Amikam Aharoni.
\newblock Demagnetizing factors for rectangular ferromagnetic prisms.
\newblock {\em Journal of applied physics}, 83:3432, 1998.

\bibitem{Ozgur2009}
Ü. Özgür, Y.~Alivov, and H.~Morkoç.
\newblock Microwave ferrites, part 1: fundamental properties.
\newblock {\em Journal of Materials Science: Materials in Electronics}, 20:789, 2009.

\bibitem{kittel2018introduction}
C~Kittel and P~McEuen.
\newblock {\em Introduction to solid state physics}.
\newblock John Wiley \& Sons, 2018.

\bibitem{cullity2011introduction}
B~Cullity and C~Graham.
\newblock {\em Introduction to magnetic materials}.
\newblock John Wiley \& Sons, 2011.

\end{thebibliography}

\end{document}